\documentclass[12pt,a4paper]{article}
\usepackage{graphicx}
\usepackage{hyperref}
\usepackage{amsmath}
\usepackage{color}
\font\Sets=msbm10
\def\Real{\hbox{\Sets R}}

\def\bb{\begin{equation}}
\def\ee{\end{equation}}
\def\pt{\partial}
\def\mod{\hbox{mod}}
\def\const{\hbox{const}}
\def\sgn{\hbox{sgn}}

\def\ve{\varepsilon}

\newtheorem{theorem}{Theorem}

\title{The capture into parametric autoresonance\footnote{This work was supported by grants RFBR 03-01-00716,
Leading Scientific Schools 1446.2003.1 and INTAS 03-51-4286.}}
\author{O.M.Kiselev,
\footnote{Institute of Mathematics of USC RAS, ok@ufanet.ru} \and 
S.G. Glebov\footnote{Ufa State Petroleum Technical University, sg@anrb.ru} }

\date{\null}
\begin{document}
\maketitle
\begin{abstract}
In this work we show that the capture into parametric resonance may be explained as the pitchfork bifurcation in the primary parametric resonance equation. We prove that the solution close to the moment of the capture is described by the Painleve-2 equation. We obtain the connection formulas for the asymptotic solution of the primary parametric resonance equation before and after the capture using the matching of the asymptotic expansions. 
\end{abstract}
\par
{\Large \bf Introduction}

\par
We study the primary parametric resonance equation:
\begin{equation}
i\epsilon\phi'+(\sigma(\theta)+|\phi|^2)\phi -\phi^*=0. \label{main-po-mu}
\end{equation}
Our goal is to develop an asymptotic theory of the capture into the parametric autoresonance with  $\sigma(\theta)\equiv-\theta$. We give the  asymptotic analysis of the capture into the parametric autoresonance and connect the asymptotic formulas for the solution of (\ref{main-po-mu}) before and after the capture.  
\par
The capture in a nonlinear (not parametric) resonance was studied in \cite{Chirikov}. The capture in the nonlinear resonance associated with loss of the stability and slow crossing through the separatrix (see \cite{Neishtadt1,Itin-Neishtadt-Vasiliev}). Autoresonance phenomenon was supposed for accelerators of relativistic particles in \cite{Veksler,McMillan}. Later the autoresonance was founded in many different oscillatory and wave processes (see review \cite{Friedland}). Mathematical approach to the capture into the autoresonance was considered in \cite{Kalyakin}. The capture into autoresonance accompanies with an bifurcation or separatrix crossing in the primary resonance equation (see \cite{Kiselev-Glebov}). 
\par
It is well-known the parametric resonance leads to the exponential grows for the solutions of  the linear equation (see \cite{Floquet}).  However the nonlinear terms lead to the unbalancing of the frequency of the external force and the frequency of the oscillations for the solution of the nonlinear equation. It changes the amplitude of the oscillations for the value of the order of square root of the perturbations.  This phenomenon  was shown  in  \cite{Bogolyubov-Mitropolskii} by  analysis of the solution for primary parametric resonance equation in the form:
\bb
R'+R\sin(2\nu)=0,\quad \nu'-\sigma-R^2+\cos(2\nu) =0,
\label{standardPrimaryParametricResonsnceEquation}
\ee
where $\sigma=\const$.
\par
Later equation  (\ref{standardPrimaryParametricResonsnceEquation}) was studied  in \cite{Fajans-Gilson-Friedland,Khain-Meerson,Asaf-Meerson}, when  $\sigma\not\equiv\const$.  It was shown that the autoresonance phenomenon takes  place for the primary parametric resonance equation when $\sigma<0$ and the modulo of the  coefficient $|\sigma|$ grows with respect to $\theta$.
\par
We consider the primary parametric resonance equation in the form (\ref{main-po-mu}). One can obtain this equation after the substitution:
$\phi=R\exp(i\nu)$. There are two reasons to investigate equation (\ref{main-po-mu}) with  $\sigma(\theta)\equiv-\theta$. On the one hand the case  $\sigma(\theta)\equiv-\theta$ is the simplest one and it contains the autoresonance phenomenon for parametrically driven systems. On the other hand this case saves  most of essential features of the solution for equation (\ref{main-po-mu}).  
\par 
Our analysis gives opportunity for an another point of view on the subject. The capture into the parametric autoresonance may be considered as a loss of stability in the pitchfork bifurcation. The dynamic theory of the pitchfork   bifurcation for ordinary second order differential equations with  slowly varying coefficient was considered in \cite{Haberman1,Maree,Haberman2}. The bifurcation when two centers and saddle coalesce is called by supercritical pitchfork. The solution in bifurcation layer approximates by solutions of the Painleve-2 equation (see \cite{Haberman1}). Later the connection formulas was obtained for the solution before and after the supercritical pitchfork bifurcation for a asymptotic solution of the special form of perturbed Painleve-2 equation with dissipation and slowly varying bifurcation parameter (see \cite{Maree}).  
\par
The supercritical pitchfork bifurcation for  equation 
$$
A'=i(\lambda A-|A|^2-\delta A^*)
$$
was considered in \cite{Guckenheimer-Mahalov}. Here $\lambda$ and $\delta$ are two parameters. We study the same equation but the parameter $\lambda$ varies slowly. 
\par
In this work we show that the capture into parametric resonance may be explained as the pitchfork bifurcation in the primary parametric resonance equation. We prove that the solution close to the moment of the capture is described by the Painleve-2 equation and obtain the connection formulas for the solution of the primary parametric resonance equation before and after the capture using the matching of the asymptotic expansions. 
\par
The contents of the paper is as follows.  The first section contains the discussions based on the numerical and qualitative analysis of the problem.  This section contains the short description of the main result also. The WKB-asymptotic expansion for the solution of (\ref{main-po-mu}) before the caption is obtained in section 2. In section 3 we match  the WKB-asymptotic expansion and the asymptotic expansion based on the increasing solution of the Painleve-2 equation.  This equation is used into a small resonant layer. Our analysis of the solution for perturbed Painleve-2 equation is based on the connection formulas for the Painleve-2 transcendent (see \cite{Its-Kapaev,Belogrudov}). The WKB-asymptotic expansion for the captured solution of (\ref{main-po-mu}) is constructed and matched with the increasing solution of the Painleve-2 equation in section 4.

\section{Discussion of the problem and main result}
\par
In this section we discuss the capture into the parametric resonance for the solution of (\ref{main-po-mu}) and present the numeric and qualitative analysis of the capture.  We formulate the main result of the work at the end of this section.
\subsection{Numeric analysis}
\par
\begin{figure}[th]{
\includegraphics{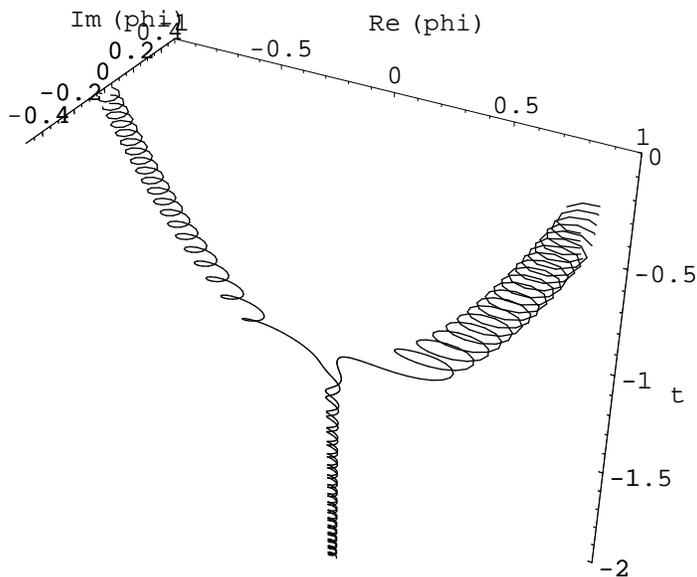}
\caption{the figure shows two solution of (\ref{main-po-mu}). Left curve shows the solution of the Cauchy problem for the equation (\ref{main-po-mu}) at $t=-2.01$, $\Re(\phi)=0.02,\,\,\Im(\phi)=0$, $\ve=0.01$. Right curve shows the solution of the Cauchy problem for the equation (\ref{main-po-mu}) at $t=-2$, $\Re(\phi)=0.02,\,\,\Im(\phi)=0$, $\ve=0.01$.}\label{numericSolutionsShowTheCapture}}
\end{figure}

Let us consider two numeric solutions for equation (\ref{main-po-mu}) (see figure \ref{numericSolutionsShowTheCapture}). These solutions differ at the initial moment only.  The first solution (left) corresponds to the initial moment at $t=-2$ and the second one (right) corresponds to $t=-2.01$.  We see the solution of the equation is very sensitive with respect to an initial data. 
\par
\begin{figure}[h]{
\includegraphics{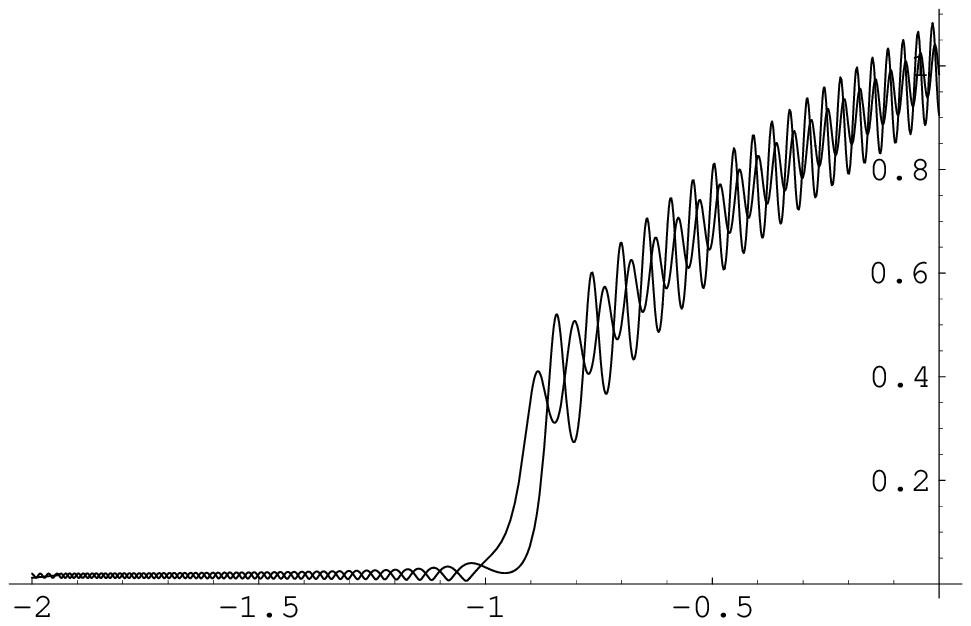}
\caption{This figure shows $|\phi|^2$ for the solutions of the Cauchy problem for equation (\ref{main-po-mu}) at $t=-2$, $\Re(\phi)=0.02,\,\,\Im(\phi)=0$, $\ve=0.01$ and at $t=-2.01$, $\Re(\phi)=0.02,\,\,\Im(\phi)=0$, $\ve=0.01$.}\label{EnergyOfTheCapture}}
\end{figure}

The picture of $|\phi|^2$ for the oscillations looks like the figure \ref{EnergyOfTheCapture} for both solutions.
\par
Figure \ref{EnergyOfTheCapture} shows the moment of the capture into the autoresonance. This moment is closed to $t=-1$.
\par
The illustrated phenomenon is the scattering on the parametric autoresonance. Our goal is to obtain the solution for the scattering problem in terms of the asymptotic theory.

\subsection{Qualitative analysis}
Here we present the qualitative analysis for the equation with the "frozen" coefficient $\sigma\equiv -T$:
\begin{equation}
i\epsilon\phi'+(-T+|\phi|^2)\phi -\phi^*=0. \label{frozen}
\end{equation}
\par
The trajectory of the solution for the equation with the varying coefficient $\sigma(\theta) =-\theta$ is locally closed to the trajectories of the equation (\ref{frozen}) at $T=\theta$.  Therefore the solution of the equation with the frozen coefficient gives a local behavior for the solution of (\ref{main-po-mu}). 
\par
The Hamiltonian for the equation is:
$$
H(\phi,\phi^*) =-{1\over2}|\phi|^4+T|\phi|^2 +{1\over2}((\phi^*)^2+\phi^2).
$$
It easy to see that there exists only one center when $T<-1$. There exist two centers and one saddle when $-1<T<1$ and three  centers and two saddles at $T>1$.  We show the phase portraits for the equation with different value of the parameter $T$ on the following  figure.
\begin{figure}[th]{
\includegraphics[width=4cm]{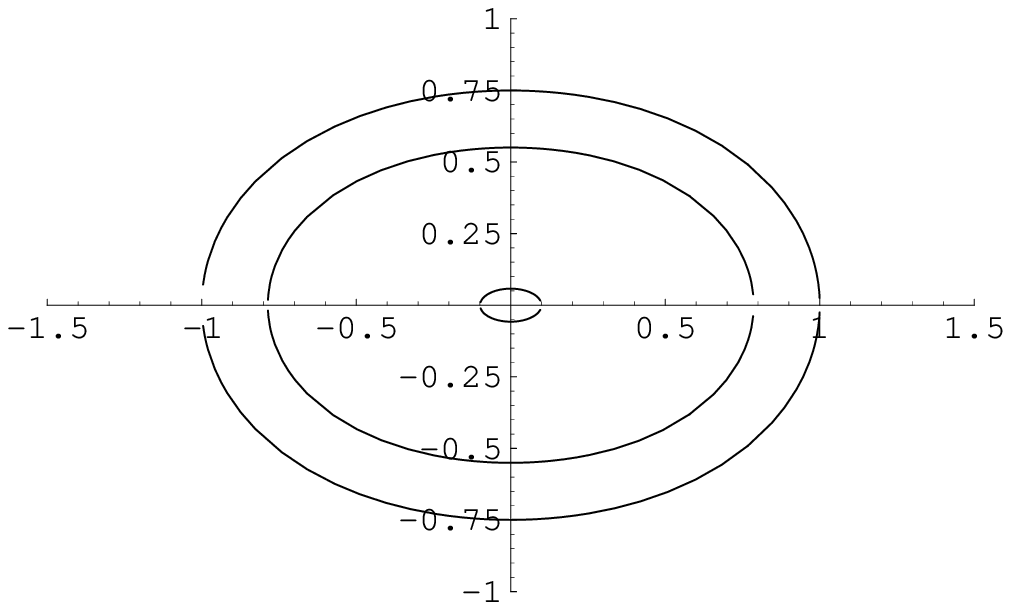} \includegraphics[width=4cm]{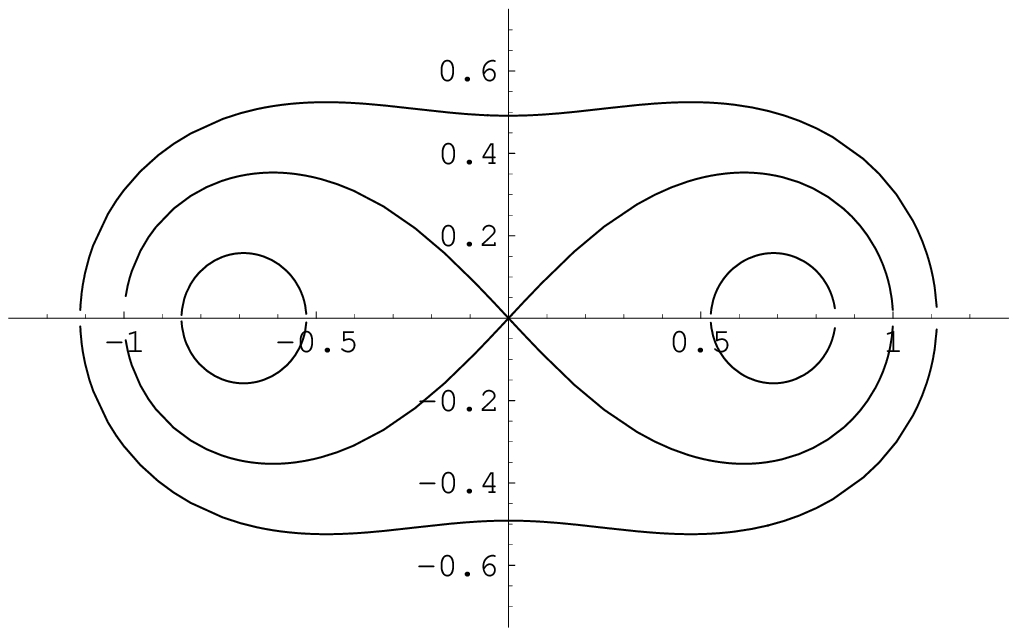}\includegraphics[width=4cm]{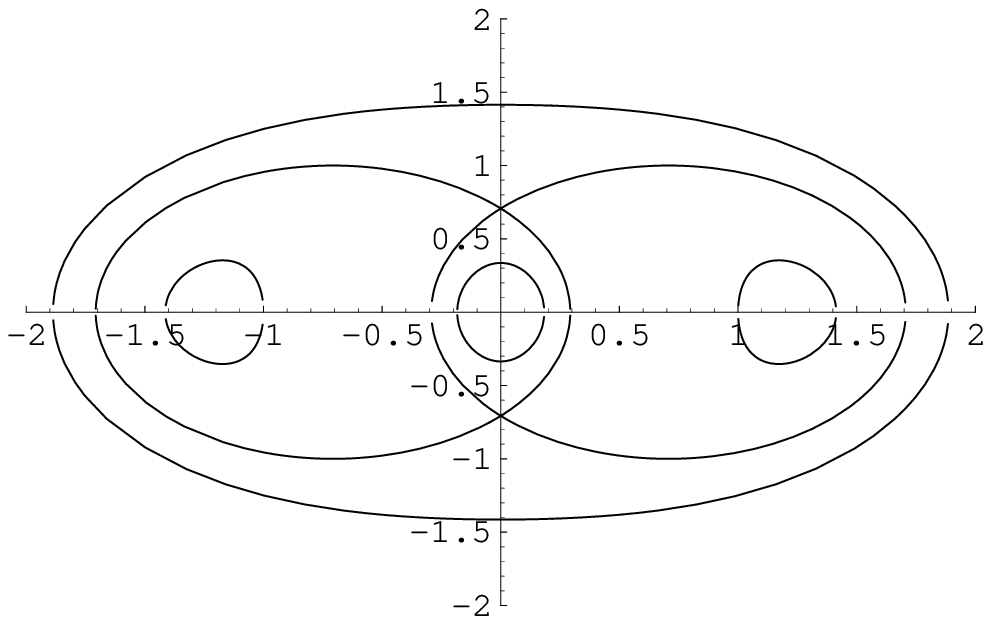}
\caption{On the left picture T < -1, on the middle picture T=0, and on the right picture T > 1.}\label{qualitativeFigure}}
\end{figure}
\par
These three pictures give the  conjecture about the numeric solution that was shown on the figures \ref{numericSolutionsShowTheCapture} and \ref{EnergyOfTheCapture}. On figure \ref{numericSolutionsShowTheCapture} one can see the non resonant solution when $t\le-1$, the capture  when $-1.1\le t\le-0.8$ and the captured solution when $-1< t$. Before the capture  the solution oscillates around the unique center for the $T<-1$. When $T > -1$ we see the captured solution that oscillates around one of two centers from the second picture in figure \ref{qualitativeFigure}. The left trajectory is captured by the left center from the second picture on figure \ref{qualitativeFigure} and the right trajectory on  figure \ref{numericSolutionsShowTheCapture} is captured by the right center on the second picture of figure \ref{qualitativeFigure}.  The bifurcation at the $T=1$ does not effect on the captured solutions. These solutions remain close to the same centers. Below we obtain asymptotic solutions that illustrate the qualitative and numeric analysis.  More over we obtain the formula that connects  the asymptotic solutions before the capture and the solutions captured by  the left  or the right center. The formula is obtained by the matching method \cite{Ilin}. 

\subsection{Main result}
Let us formulate the main result.
\begin{theorem}\label{mainStatement}
Let the asymptotic solution of the primary  resonance equation be
$$
\phi\sim\ve^{1/2}\alpha_{1,0}\bigg[\sqrt[4]{\theta-1\over\theta+1}\sin\big(s\big)+ i\sqrt[4]{\theta+1\over\theta-1}\cos\big(s\big)\bigg],\quad \hbox{as}\ \theta<-1,
$$
where 
$$
s={\omega\over \ve}+\varphi_{1,0}+\alpha_{1,0}^2\bigg[2\theta+2\ln\bigg|{\theta-1\over \theta+1}\bigg|\bigg],
$$
and
$$
\omega={1\over2}\theta\sqrt{\theta^2-1}-{1\over2}\ln(\theta+\sqrt{\theta^2-1}).
$$
The constants $\alpha_{1,0}$ and $\varphi_{1,0}$ are the parameters of the solution and 
$$
\varphi_{1,0}\not={3\over2}\alpha_{1,0}^2\ln(2)-{\pi\over4}-\arg\bigg(\Gamma(i\alpha_{1,0}^2/2)\bigg) +\chi\pi(\mod(2\pi)),\chi=0,1, 
$$
then the asymptotic solution  has a form:
\begin{eqnarray}
\Phi^{(j)}(S,\theta,\ve)\sim (-1)^j\bigg[{1\over2}\sqrt{1+\theta} + \ve^{1/2} A_{0,0}\bigg({1\over\sqrt[4]{1+\theta}}\cos(S) + i\sqrt[4]{1+\theta}\sin(S)\bigg)\bigg],
\label{mainTermOfCapturedAsymptotics}
\\
S\sim{\Omega\over\ve}+\varphi_{0,0}+A_{0,0}^2\bigg({5\over2}\theta+{3\over4}\theta^2+{1\over6}(112-8\theta)\sqrt{1+\theta}+{3\over2}\ln(1+\theta)\bigg)\nonumber
\end{eqnarray}
in the domain $\theta>-1$.
Here 
$$
\Omega\sim{4\over3}(\theta+1)^{3/2},
$$
$j=2,3$, the constants $\varphi_{0,0}$ and $A_{0,0}$ are the parameters of the asymptotic solution. The parameters are defined by 
\begin{eqnarray*}
A_{0,0}^2={1\over\pi}\ln\bigg({1+|p|^2\over 3|\Im(p)|}\bigg),\\
\varphi_{0,0}=-{\pi\over4}+{7\over2}A_{0,0}^2-\arg\bigg(\Gamma(iA_{0,0}^2)\bigg)-\arg(1+p^2),
\end{eqnarray*}
where
$$
p=\sqrt{\exp(\pi \alpha_{1,0}^2)-1} \exp\bigg(i{3\over2}\alpha_{1,0}^2\ln(2)-i{\pi\over4}- i\arg\big(\Gamma(i\alpha_{1,0}^2/2)\big)-i\varphi_{1,0}\bigg).
$$
In formula (\ref{mainTermOfCapturedAsymptotics}) $j=2$ as $\Im(p)>0$ and $j=3$ as $\Im(p)<0$.
\end{theorem}

\section{WKB-solution before the capture}
\label{sectionWKBasymptotcSoutionBeoforeTheCapture}
\par
Let us construct the solution of the WKB-type before the capture. The qualitative analysis shows that the capture into the parametric resonance occurs at $\theta=-1$. When $\theta<-1$ the solution of  equation (\ref{frozen}) has an unique center at $\psi=0$. This fact prompts that equation (\ref{main-po-mu}) has the oscillating  solutions and they are closed to $\psi=0$. We construct the asymptotic expansion for the solutions of such type in this section.

\subsection{The WKB solution closed to zero}
\par
Let us consider the WKB-solution in the form:
\bb
\phi=\ve^{1/2}\sum_{n=1}^\infty \ve^{n-1} \phi_n(s,\theta).
\label{smallWKBsolution}
\ee
The leading-order term of the asymptotic expansion is the solution of the equation:
\bb
i \omega' \pt_s\phi_1-\theta\phi_1-\phi_1^*=0.
\label{linearEquationNearZero}
\ee
\par
The higher-order terms of (\ref{smallWKBsolution}) are solutions of the equation:
\bb
i \omega \phi_n'-\theta\phi_n-\phi_n^*=f_n, \quad n=2,3,\dots,
\label{linearEquationNearZeroForHigherOrderTerms}
\ee
where
$$
f_n=-i\pt_\theta \phi_{n-1} -i\pt_\theta\varphi_{n-1}\pt_s\phi_n +h_n.
$$
Here $h_n$ is a polynomial of the third order  with respect to  $\phi_m$, $m<n$.

\paragraph{The basis of the solutions for the linear equation}
\par
To solve  equation (\ref{linearEquationNearZero}) and the equations for the higher-order terms we construct the basis of two linear independent solutions for equation (\ref{linearEquationNearZero}) as the WKB-approximation. Define the basis solutions for equation (\ref{linearEquationNearZero}) by $v_1$ and $v_2$. Let the first solution be
$$
v_1(\tau,\theta)=\omega'\cos\big(\tau\big) - i (\theta+1)\sin\big(\tau\big),
$$
where, $\tau=\omega/\ve$,
$$
(\omega')^2=\theta^2-1.
$$
The Wronskian of two linear independent solutions of equation (\ref{linearEquationNearZero}) is constant. Therefore we construct the second solution as:
$$
v_2(\tau,\theta)={-1\over (\theta+1)}\sin\big(\tau\big) - {i\over \omega'}\cos\big(\tau\big).
$$
The Wronskian for these solution is:
$$
w(v_1,v_2)=v_1 v_2^*-v_1^*v_2=2i.
$$
The formula for the general solution of (\ref{linearEquationNearZero}) has a form:
$$
v(\tau,\theta)=C_1(\theta)v_1(\theta,S)+C_2(\theta)v_2(\tau,\theta).
$$
\par
Using the WKB approximation we specify the $C_1(\theta)$ and $C_2(\theta)$ such that:
$$
C_1=C_{1,0}{1\over\sqrt{({-\theta-1})\omega'}}.
$$
$$
C_2=C_{2,0}\sqrt{(-\theta-1)\omega'}.
$$
Here $C_{1,0}$ and $C_{2,0}$ are arbitrary constants. 
\par
We use the functions
$$
w_1(\tau,\theta)=\bigg[\bigg({\theta-1\over \theta+1}\bigg)^{1\over4}\cos(\tau)+ i \bigg({\theta+1\over \theta-1}\bigg)^{1\over4}\sin(\tau)\bigg]
$$
and 
$$
w_2(\tau,\theta)=\bigg[\bigg({{\theta-1}\over{\theta+1}}\bigg)^{1\over4}\sin(\tau)- i \bigg({\theta+1\over \theta -1}\bigg)^{1\over4}\cos(\tau)\bigg]
$$
as the basis of the linear independent solutions of equation (\ref{linearEquationNearZero}).

\subsection{Constructing of the WKB solution for nonlinear equation (\ref{main-po-mu})}
\par 
The main term of the WKB asymptotic solution for (\ref{main-po-mu}) we construct in the form:
\bb
\phi_1(s, \theta)=\alpha\bigg[\bigg({{\theta-1}\over{\theta+1}}\bigg)^{1\over4}\sin(s)- i \bigg({{\theta+1}\over{\theta-1}}\bigg)^{1\over4}\cos(s)\bigg],
\label{WKBSolutionAsPhiTo0}
\ee
where $s=\omega/\ve+\varphi$ and arbitrary constants $\alpha$ and  $\varphi$ are parameters of the solution. These parameters are represented by asymptotic series:
$$
\alpha=\sum_{k=1}^\infty \ve^{k-1}\alpha_k(\theta),\quad \varphi=\sum_{k=1}^\infty \ve^{k-1}\varphi_k(\theta).
$$
\par
It is easy to see that  the solution of (\ref{linearEquationNearZeroForHigherOrderTerms}) is bounded if the right-hand side is orthogonal for two linear independent solutions of the linearized equation.   
\begin{eqnarray}
\int_{0}^{2\pi} f_n(z,\theta)w_1^*(z,\theta)+f_n^*(z,\theta)w_1(z,\theta) dz=0,\nonumber
\\
\label{orthogonalityFormulas}
\\
\int_{0}^{2\pi} f_n(z,\theta)w_2^*(z,\theta)+f_n^*(z,\theta)w_2(z,\theta) dz=0.\nonumber
\end{eqnarray}
\par
Let us construct the bounded solution of (\ref{linearEquationNearZeroForHigherOrderTerms}) at $n=2$. The right-hand side is 
\begin{eqnarray*}
f_2=-i{{d}\over{d\theta}}\phi_1 -|\phi_1|^2\phi_1\equiv - i\pt_\theta \alpha_1\pt_\alpha\phi_1 - i \alpha_1\pt_s\phi_1 - |\phi_1|^2\phi_1.
\end{eqnarray*}
Formulas (\ref{orthogonalityFormulas}) give the equations for the functions $\alpha_1(\theta)$ and $\varphi_1(\theta)$:
$$
\pt_\theta \alpha_1=0,\quad 2\pt_\theta\varphi_1 +\alpha_1^2|\phi_1|^2=0.
$$
Using the formula for $\phi_1$ we obtain: 
$$
2\pt_\theta\varphi_1 +\alpha_1^2{(\theta+1)^2+(\theta-1)^2\over (\theta+1)(\theta-1)}=0.
$$
The solution of the equation has the form:
$$
\varphi_1 =\varphi_{1,0} - (\alpha_1^0)^2\bigg(\theta + \ln\bigg|{\theta-1\over\theta+1}\bigg|\bigg).
$$
\par
Analogously, the next terms of the WKB-asymptotic expansion are constructed. The equation for the $n$-th correction term of the WKB-asymptotic solution has a form:
\bb
i \omega' \phi_n'-\theta\phi_n-\phi_n^*=f_n,
\label{linearEquationNearZeroForNthCorrection}
\ee
where
$$
f_n=\pt_\theta\alpha_{n-1}\pt_\alpha\phi_1+\alpha \pt_\theta\varphi_{n-1}\pt_s\phi_1+h_n.
$$
Here $h_n$ depends on the lower order terms of the WKB asymptotic expansion.
\par
These equations allow to determine coefficients $\alpha_n$ and $\varphi_n$.

\paragraph{Asymptotic behavior close to the turning point}
\par
The WKB asymptotic expansion is not valid at the turn point $\theta=-1$. As $\theta\to-1-0$ the asymptotic behavior of the coefficients  for asymptotic series (\ref{smallWKBsolution}) are represented by following formulas. For the leading-order term we obtain
$$
\phi_1\sim\alpha_{1,0}\bigg({-2\over \theta+1}\bigg)^{-1/4}\sin\bigg({2\sqrt{2}\over3}(1-\theta)^{3/2}- (\alpha_{1,0})^2\ln(1+\theta)+\varphi_{1,0}\bigg),\ \theta\to-1-0
$$
where $\alpha_{1,0}=\const$ and $\varphi_{1,0}=\const$ are the parameters of the solution. 
\par
The second term of asymptotic series (\ref{smallWKBsolution}) has a behavior: 
$$
\phi_2=O((-1-\theta)^{-7/4})\quad \theta\to-1-0.
$$
The recurrent calculations give
$$
\phi_{n+1}=O(\phi_{n} (-1-\theta)^{-3/2})),\quad \theta\to-1-0.
$$

\paragraph{The domain of validity}
\par
Expansion (\ref{smallWKBsolution}) saves the asymptotic property with respect to $\ve$ when  
$$
{\ve\phi_{n+1}\over \phi_{n}}\ll1,\quad\hbox{as}\quad \theta\to-1-0. 
$$
It yields the domain of validity for (\ref{smallWKBsolution})
$$
\ve^{-2/3}(-1-\theta)\gg1.
$$

\section{The Painleve layer} 
\par
To construct the asymptotic solution when $\theta$ is closed to $-1$ we use the scaled variables:
$$
\theta+1=\ve^{2/3}\eta,\quad \phi=\ve^{1/3}x(\eta,\ve)+i\ve^{2/3}y(\eta,\ve).
$$
This change of the variables leads to the system of the equations for real and imaginary parts of the function $\phi$:
\begin{eqnarray}
x'+2y=\ve^{2/3}(\eta-x^2)y - \ve^{4/3}y^3,\nonumber\\
y'+(\eta-x^2)x=\ve^{2/3}y^2 x.
\label{perturbedPainleve2}
\end{eqnarray}
\subsection{The asymptotic expansion in the Painleve layer}
\par
We construct the solution of system (\ref{perturbedPainleve2}) of the form:
\bb
x(\eta,\ve)=\sum_{n=0}^\infty \ve^{2n/3}x_n(\eta),\quad y(\eta,\ve)=\sum_{n=0}^\infty \ve^{2n/3}y_n(\eta). \label{asymptoticSolutionOfPerturbedPainleveEquation}
\ee
The leading-order terms of (\ref{asymptoticSolutionOfPerturbedPainleveEquation}) 
 are solutions of the system of equations:
\begin{eqnarray}
x_0'+2y_0=0,\nonumber\\
y_0'+(\eta-x_0^2)x_0=0.\label{systemForTheMainTerm}
\end{eqnarray}
The higher-order terms are determined from the system:
\begin{eqnarray}
x_n'+2y_n=H^{(1)}_n,\nonumber\\
y_n'+(\eta-3x_0^2)x_n=H^{(1)}_n.\label{systemForTheHigherOrederTerm}
\end{eqnarray}
Here functions $H^{(1)}_n$ and $H^{(2)}_n$ depend on $\eta$ and the lower-order terms of asymptotic expansion (\ref{asymptoticSolutionOfPerturbedPainleveEquation}). For example 
$$
H_1^{(1)}=(\eta-x_0^2)y_0,\quad H_1^{(2)}=y_0^2 x_0;
$$
$$
H_2^{(1)}=(\eta-x_0^2)y_1 -2x_0y_0x_1-y_0^3,\quad H_2^{(2)}=2 y_0 y_1 x_0 + y_0^2 x_1;
$$
\par
The leading-order term of the asymptotic expansion $x_0$ is a solution of the second order equation: 
\begin{equation}
x_0''+2(-\eta+x_0^2)x_0=0.
\label{equationForThemainTermOfTheInternalAsymptotics}
\end{equation}
\par 
The higher-order terms of (\ref{asymptoticSolutionOfPerturbedPainleveEquation}) satisfy to the linear differential equation of the second order:
\bb
x_n''+2(-\eta+3x_0^2)x_n=h_n,\quad  h_n=H_n^{(2)}+2{d\over d\eta}H_n^{(1)}.\label{equationForNcorrTermInPainleveLayer}
\ee
\par
The right hand side of (\ref{equationForNcorrTermInPainleveLayer}) has the following  structure:
$$
h_n=H_n^{(2)}+2 \pt_\eta H_n^{(1)}+ 2\sum_{k=0}^{n-1}\pt_{x_k}H_n^{(1)}x_k'+2 \sum_{k=0}^{n-1}\pt_{y_k}H_n^{(1)}y_k'.
$$
The substitution $x_k'=-2y_k+H_k^{(1)}$ and $y_k'=(3x_0^2-\eta)x_k + H_k^{(2)}$ gives:
\begin{eqnarray}
h_n=H_n^{(2)}+2 \pt_\eta H_n^{(1)}+ 2\sum_{k=0}^{n-1}\pt_{x_k}H_n^{(1)}(-2y_k+H_n^{(1)}) +
\nonumber\\
 2\sum_{k=0}^{n-1}\pt_{y_k}H_n^{(1)}((3x_0^2-\eta)x_k + H_k^{(2)}).
\label{rightHandSideOfTheLinearizedEquation}
\end{eqnarray}
\par
The substitutions
\begin{equation}
\eta=2^{1/3}z,\quad x_0(\eta)=2^{1/3}iu(z)\label{substitutionForPainleve}
\end{equation}
reduce (\ref{equationForThemainTermOfTheInternalAsymptotics}) to the Painleve-2 equation:
\begin{equation}
u''-zu-2u^3=0.
\label{Painleve2}
\end{equation}
Using the substitution $x_n(\eta)=2^{1/3}iu_n(z)$ we obtain the perturbed linearized Painleve-2 equation:
\begin{equation}
u_n''-z u_n-6u^2 u_n=H_n.
\label{linearizedPainleve2}
\end{equation}
\par
The Painleve-2 equation  is integrable by the isomonodromic deformation method  \cite{Flaschka-Newell}. The solution $u(z;{\tilde\alpha},{\tilde\varphi})$  of the equation is called by the Painleve transcendent that depends on two parameters ${\tilde\alpha}$ and ${\tilde\varphi}$. It is known that the real solution of (\ref{Painleve2})  does not have the singularities as $z\in\Real$ \cite{Its-Novokshenov}. Therefore  the main term of the asymptotic expansion is presented by the Painleve transcendent $u(z,{\tilde\alpha},{\tilde\varphi})$.
\par
The homogeneous linearized Painleve-2 equation:
$$
w''-z w +6u^2w=0
$$
has two linear independent solutions:
$$
u_1=\pt_{\tilde\alpha} u(z,{\tilde\alpha},{\tilde\varphi}), \quad u_2=\pt_{\tilde\varphi} u(z,{\tilde\alpha},{\tilde\varphi}).
$$
These solutions allow to present the higher-order terms of the expansion (\ref{asymptoticSolutionOfPerturbedPainleveEquation}) in the form:
\begin{eqnarray}
x_n=B_n^{(1)} X_1(\eta)+ B_n^{(2)} X_2(z)+ X_1(z)\int_{\eta_0}^\eta {H_n(\zeta)X_2(\zeta)\over W}d\zeta -
\nonumber\\ 
X_2(\eta)\int_{\eta_0}^\eta {H_n(\zeta)X_1(\zeta)\over W}d\zeta.
\label{SolutionForN-thOrederTerm}
\end{eqnarray}
Here $W\equiv\const\not=0$ is the Wronskian of the solutions $X_1=i u_1(2^{-1/3}\eta)$ and $X_2=i u_2(2^{-1/3}\eta)$ for the linearized  equation, $B_n^{(1)}$, $B_n^{(2)}$ and $\eta_0$ are arbitrary constants.
\paragraph{The asymptotic behavior on the left-hand side of the validity interval}
\par
In this paragraph we determine the domain of validity for  asymptotic expansion (\ref{asymptoticSolutionOfPerturbedPainleveEquation}) and match the expansion with the WKB-asymptotic expansion that was obtained in section \ref{sectionWKBasymptotcSoutionBeoforeTheCapture}. Therefore we should determine the asymptotic behavior of the Painleve-2 transcendent $u(z,{\tilde\alpha},{\tilde\varphi})$ as $z\to-\infty$.
\par
The asymptotic expansion of the Painleve-2 solution as $z\to-\infty$ has a form (\cite{Its-Kapaev,Belogrudov}):
\bb
u(z)=i{\tilde\alpha}(-z)^{-1/4}\sin({2\over3}(-z)^{3/2}+{3\over4}{\tilde\alpha}^2\ln(-z)+{\tilde\varphi})+o((-z)^{-1/4}).
\label{AsymptoticAsMinusInfty}
\ee
Here  ${\tilde\alpha}$ and ${\tilde\varphi}$ are parameters of the solution of Painleve-2 equation.
\par 
Formula (\ref{AsymptoticAsMinusInfty}) give the asymptotic behavior for the solutions of the linearized equation $U_1$ and $U_2$ as $z\to-\infty$:
$$
U_1\equiv \pt_{\tilde\alpha} u\sim i{3\over2}{\tilde\alpha}^2\ln(-z)(-z)^{-1/4}\cos({2\over3}(-z)^{3/2}+{3\over4}{\tilde\alpha}^2\ln(-z)+{\tilde\varphi}),
$$
$$
U_2\equiv \pt_{\tilde\varphi} u\sim i{\tilde\alpha}(-z)^{-1/4}\cos({2\over3}(-z)^{3/2}+{3\over4}{\tilde\alpha}^2\ln(-z)+{\tilde\varphi}).
$$
\par
Using these asymptotic behaviors we obtain the asymptotic formulas for the higher-order terms as $z\to-\infty$:
\begin{equation}
x_n=A_n^- X_1(\eta)+ B_n^-X_2(\eta)+ {\mathbf X}_n^-(z).
\label{SolutionForN-thOrederTermAsMinusInfinity}
\end{equation}
Here $A_n^-$ and $B_n^-$ are constants. They are obtained by matching procedure with external asymptotic solution below and ${\mathbf X}_n^-(z)$ is the particular solution of the equation for the $n$-th term of the asymptotic expansion  $x_n$.
\paragraph{The left border of the interval of the validity for the Painleve layer}
\par
Using the asymptotic behavior of $x_0$ as $\eta\to-\infty$ and the asymptotic behavior of $X_1$ and $X_2$ we obtain the asymptotic behavior of the higher-order terms. The equation for $x_1$ is
$$
x_1''+(-\eta+x_0^2)x_1=O((-\eta)^{2-1/4}), \quad \eta\to-\infty.
$$
The particular solution has an order:
$$
{\mathbf X}_1^-=O((-\eta)^{3/4}), \quad \eta\to-\infty.
$$
The higher-order term satisfies the equation of the form
$$
x_n''+(-\eta+x_0^2)x_n=O((-\eta)^{2}x_{n-1}).
$$
By the sequential calculations we obtain
$$
{\mathbf X}_n^-=O((-\eta)^{n-1/4}).
$$
\par
This estimated value for the higher-order terms give the left border for the interval of validity for (\ref{asymptoticSolutionOfPerturbedPainleveEquation}):
$$
{\ve^{2(n+1)/3}x_{n+1}\over \ve^{2n/3}x_n}\sim{\ve^{2/3}X_{n+1}^-\over X_n^-}\ll1,\quad \ve^{2/3}(-\eta)\ll1,
$$
or
$$
-\ve^{2/3}\eta\ll1,\quad \eta<0,\,\,\,\ve\to0.
$$

\subsection{Matching with the WKB asymptotic expansion}
\par
The domains of validity of  asymptotic expansions (\ref{smallWKBsolution}) and (\ref{asymptoticSolutionOfPerturbedPainleveEquation}) are intersected. The matching with the WKB asymptotic expansion from section \ref{subsectionWKBasymptotics} allows to obtain the parameters ${\tilde\alpha}$ and $\tilde\varphi$ for the solution of the Painleve-2 equation, for example:
\bb
{\tilde\alpha}=\alpha_{1,0},\quad {\tilde\varphi}=\varphi_{1,0}.\label{matchingBeforeTheCapture}
\ee

\paragraph{Asymptotic behavior on the right-hand side}
\par
The   theorem by Its-Kapaev-Belogrudov \cite{Its-Kapaev,Belogrudov} gives the asymptotic behavior of  solution (\ref{AsymptoticAsMinusInfty}) for the Painleve-2 equation. 
\begin{theorem}[by Its-Kapaev-Belogrudov]\label{theoremByItsKapaevBelogrudov}
If  a solution of the Painleve 2 equation has asymptotic behavior (\ref{AsymptoticAsMinusInfty}) as $z\to -\infty$ and 
$$
{\tilde\varphi}_1={3\over2}{\tilde\alpha}^2\ln(2)-{\pi\over4}-\arg\bigg(\Gamma\bigg({i{\tilde\alpha}^2\over2}\bigg)\bigg)+\kappa\pi(\mod(2\pi), \quad \kappa=0,1,
$$
then the solution of the Painleve-2 equation is:
$$
u(z)={iq\over2\sqrt{\pi}}z^{-1/4}\exp\bigg(-{2\over3}z^{3/2}\bigg)(1+o(1)),\quad z\to\infty,
$$
where $a^2=\exp(\pi {\tilde\alpha}^2) - 1$, and $\sgn(a)=1-\kappa/2$;
\newline
Otherwise  the absolute value of the solution  increases:
\bb
u(z)=\pm i\sqrt{{z\over2}}\pm i(2z)^{-1\over4}\rho\cos\bigg({2\sqrt{2}\over3}z^{3/2}-{3\over2}\rho^2\ln(z)+\upsilon\bigg) + o(z^{-1/4}),
\label{asymptoticsOfThePainleveTranscendentAtPlusInfinity1}
\ee
where $\rho>0$ and $0\le\upsilon<2\pi$ .  The parameters $\rho$ and $\upsilon$ is defined by ${\tilde\alpha}$ and $\tilde\varphi$:
$$
\rho^2={1\over\pi}\ln\bigg({1+|p|^2\over3|\Im(p)|}\bigg),
$$
$$
\upsilon=-{\pi\over4}+{7\over2}\rho^2\ln(2)-\arg(\Gamma(i\rho^2)-\arg(1+p^2),
$$
where
$$
p=(\exp(\pi{\tilde\alpha}^2-1)^{1/2}\exp\bigg(i{3\over2}{\tilde\alpha}^2\ln(2)-i{\pi\over4}-i\arg(\Gamma(i{{\tilde\alpha}^2\over2}))-i\tilde\varphi\bigg),
$$
and the sign "+" is defined by $\Im(p)<0$;
\newline
if $\rho=0$, then
\bb
u(z)=\pm i\sqrt{{z\over2}}\pm i{z^{-5/2}\over 8\sqrt{2}} + O(z^{-11/2}).
\label{asymptoticsOfThePainleveTranscendentAtPlusInfinity2}
\ee
\end{theorem}
These formulas give the asymptotic solution of the system of equations (\ref{perturbedPainleve2}) as $\eta\to\infty$. The solution of the scattering problem for the system of equations  (\ref{perturbedPainleve2}) allows to solve the problem of the capture into the resonance for the small amplitude solution of the primary resonance equation. 

\paragraph{The right border of the interval of the validity for the asymptotic expansions in the Painleve layer}
\par
The formulas (\ref{asymptoticsOfThePainleveTranscendentAtPlusInfinity1}), (\ref{asymptoticsOfThePainleveTranscendentAtPlusInfinity2})and substitutions (\ref{substitutionForPainleve}) give the leading-order term of the captured into the resonance asymptotic solution. The solution has the behavior:
$$
x_0(\eta)\sim\mp 2^{1/3}\sqrt{\eta},\quad \eta\to\infty.
$$
The nonlinear terms in the perturbed Painleve-2 equation (\ref{perturbedPainleve2}) are 
$$
x_n(\eta)=O((-\eta)^{n+1/2}).
$$
Then asymptotic expansion (\ref{asymptoticSolutionOfPerturbedPainleveEquation}) is valid in the domain 
$$
\ve^{2/3}\eta\ll1.
$$

\section{The captured WKB-asymptotic solution}\label{sectionTheCapturedWKB}
\par
The previous section shows the asymptotic solution of  equation (\ref{main-po-mu}) increases as $\pm\sqrt{\theta}$. When $\theta\to -1+0 $ the leading-order term of the asymptotic expansion in the Painleve layer corresponds to the one of the centers for the equation with frozen coefficient. Equation (\ref{main-po-mu}) has two slowly varying solutions as $-1<\theta<1$. In this section we construct the formal asymptotic expansion for these solutions and for slowly varying solutions as $\theta>1$ also. 

\subsection{Slowly varying solutions}
\label{subSecSlowlyVaryingSolution}
\par
Let us to construct the solution of the equation (\ref{main-po-mu}) as follows:
\bb
U(\theta,\ve) = \sum_{n=0}^\infty \ve^n U_n(\theta)\label{asymptoticExpansionForSlowlyVaryingSolition}
\ee
\par
The main term of the asymptotic expansion is defined by an algebraic equation:
\begin{equation}
-\theta U_0 +|U_0|^2 U_0 - U_0^*=0.
\label{algebraicEquationForMainTerm}
\end{equation}
Using the complex conjugated function we obtain:
$$
[U_0+U_0{}^*][U_0 -U_0{}^*]=0
$$
This formula shows that the leading-order term of the asymptotic expansion is pure real or pure imaginary one. The real terms are 
\bb
U_0^{(1)}=0,\quad U_0^{(2,3)} = \pm \sqrt{1+\theta},
\label{mainTermOfCenters}
\ee
and the imaginary terms are:
\bb
U_0^{(4,5)} = \pm i\sqrt{\theta-1}.\label{mainTermOfSaddle}
\ee
The algebraic equations for the higher-order terms $U_n^{(j)},\,\,\, j=2,3,4,5$ are follows:
$$
[-\theta+2|U_0^{(j)}|^2]U_n^{(j)} + [(U_0^{(j)})^2-1](U_0^{(j)}){}^* =Q_n.
$$
Here
$$
Q_n=-i(U_0^{(j)})'+\sum_{k+l+m=n}C_{klm}U_k^{(j)}U_l^{(j)}(U_m^{(j)}){}^*,
$$
where $C_{klm}$ is a constant.
\par
Let us represent the linear equation as a system of the equations for the real and imaginary parts. The determinant of the system of equations is:
$$
\Delta^{(j)}= 4(1+\theta),\quad j=2,3
$$
and 
$$
\Delta^{(j)}=4(1-\theta),\quad j=4,5.
$$
Then the system for the higher-order terms is solvable if $j=2,3$ and $\theta>-1$. Otherwise if $j=4,5$, then the system for higher-order terms is solvable when $\theta>1$.
\par
When $\theta\to\infty$ the order of $n$-th term of asymptotic is $O(\theta^{-1/2})$, as $n=1,2,\dots$. Therefore the asymptotic expansion (\ref{asymptoticExpansionForSlowlyVaryingSolition}) is uniform with respect to  $\theta$ when $\theta>-1$ for  $j=2,3$ and when $\theta>1$ for $j=4,5$.

\subsection{WKB-asymptotic expansion close to the slowly varying centers}\label{subsectionWKBasymptotics}
\par
Here we construct the WKB-asymptotic expansion that oscillates close to the slowly varying centers $U^{(j)}$. The WKB asymptotic expansion has a following form:
\begin{equation}
\Phi(\theta,\ve)=U^{(j)}(\theta,\ve)+\sum_{k=1}^{\infty}\ve^{k/2}\Phi_k^{(j)}(S,\theta,\ve).
\label{increasingAsymptoticSolution}
\end{equation}
Here 
$$
S=\Omega(\theta)/\ve+\sum_{k=0}^\infty\ve^{k/2}\varphi_k(\theta),
$$
The equations for the higher-order terms are:
\begin{equation}
i\Omega'\pt_S\Phi_k^{(j)}{}+[-\theta+2|U_0^{(j)}|^2]\Phi_k^{(j)} + [(U_0^{(j)})^2 - 1]\Phi_k^{(j)}{}^*=G_n,\label{linrzdeq}
\end{equation}
$$
G_n=-i\sum_{l+k=n-2}\pt_S \Phi_l^{(j)}\varphi_l' -i\pt_\theta\Phi_{n-2}^{(j)} -\sum_{k+l+m=n}C_{klm} (U_0^{(j)})\Phi_k^{(j)} \Phi_l^{(j)} \Phi_m^{(j)}{}^*,
$$
where $C_{klm}(U_0^{(j)})$ are the polynomials.

\paragraph{WKB-solutions for the equation in the variations}
\par
The homogeneous linearized equation of (\ref{linrzdeq}) is called by equation in the variations: 
$$
i\Omega'\pt_{\theta_1}V+(-\theta +2|U^{(j)}|^2)V +[U^{(j)}{}^2 - 1]V^*=0.
$$
Two linear independent solutions of this equation have the following forms. The first  asymptotic solution is: 
$$
V_1(\theta_1,\theta)=\Omega'\cos(\theta_1)+i\bigg((-\theta +2|U^{(j)}|^2)+ [U^{(j)}{}^2- 1]\bigg)\sin(\theta_1).
$$
Here $\theta_1=\Omega/\ve$,
$$
\Omega'=\sqrt{\big[-\theta+2|U^{(j)}|^2\big]^2 - \big|U^{(j)}{}^2-1\big|^2},
$$
The second asymptotic solution is:
\begin{eqnarray*}
V_2(\theta_1,\theta)={1\over \big[-\theta+2|U^{(j)}|^2\big] + \Re(U^{(j)}{}^2-1)}\sin(\theta_1) - 
\\
i{(U^{(j)}{}^2-1)+(-\theta+2|U^{(j)}|^2\over \Omega'(\big[-\theta+2|U^{(j)}|^2\big] + \Re(U^{(j)}{}^2-1))}\cos(\theta_1).
\end{eqnarray*}
The Wronskian of these solutions is:
$$
W(V_1,V_2)=V_1V_2^*-V_1^*V_2=2i.
$$
\par
Using explicit formulas for the asymptotic expansion of $U^{(j)} for j=2,3$ as $\ve\to0$ one can obtain:
\begin{eqnarray*}
\Omega'\sim2\sqrt{1+\theta},\\
V_1\sim 2\sqrt{1+\theta}\cos(\theta_1)+2i(\theta+1)\sin(\theta_1)\\
V_2\sim {1\over 2(1+\theta)}\sin(\theta_1)-i{1\over2\sqrt{\theta+1}}\cos(\theta_1).
\end{eqnarray*}

\paragraph{The first correction term of the WKB asymptotic expansion}
\par
Let the first correction term of the WKB-asymptotic solution $\Phi_1^{(j)}(S,\theta,\ve)$ be 
$$
\Phi_1^{(j)}(S,\theta,\ve)=A(\theta,\ve)V_1(S,\theta),
$$
where
$$
A(\theta,\ve)=\sum_{k=0}^\infty\ve^{k/2} A_k(\theta).
$$
\par
The first correction term of the asymptotic expansion (\ref{increasingAsymptoticSolution}) has two series of parameters $A_k(\theta)$ and $\varphi_k(\theta)$, $k\in {\mathbf N}$. These parameters are defined by usual way in the WKB theory. The parameters are solutions of the anti resonant equations:
\begin{eqnarray}
\int_{0}^{2\pi} G_k(z,\theta)V_1^*(z,\theta)+G_k^*(z,\theta)V_1(z,\theta) dz=0,\nonumber
\\
\label{orthogonalityFormulasPostResonantLayer}
\\
\int_{0}^{2\pi} G_k(z,\theta)V_2^*(z,\theta)+G_k^*(z\theta)V_2(z,\theta) dz=0.\nonumber
\end{eqnarray}
The pair of the equations are defined parameters $A_{k-1}(\theta)$ and $\varphi_{k-3}(\theta)$, where $k>2$.
\par
Below we obviously show the process for definition of the parameters  for $A_1(\theta)$ and $\varphi_0(\theta)$.

\paragraph{The second-order correction term of asymptotic expansion (\ref{increasingAsymptoticSolution}).}
The substitution of $\Phi_1^{(j)}$ in the   WKB asymptotic expansion leads to the equation for the second correction term:
\begin{eqnarray*}
i\Omega'\pt_S\Phi_2^{(j)}+\bigg(-\theta+2|U^{(j)}|^2\bigg)\Phi_2^{(j)}+\bigg(U^{(j)}{}^2-1\bigg)(\Phi_2^{(j)})^*=G_2,\\
G_2=-\bigg[2|\Phi_1^{(j)}|^2 U^{(j)} + (\Phi_1^{(j)})^2 (U^{(j)})^*\bigg]
\end{eqnarray*}
\par
The particular solution of the inhomogeneous equation has a form:
\begin{eqnarray*}
\tilde\Phi_2^{(j)}=V_1(S,\theta)\int^Sd{\sigma\over-2}\bigg(G_2(\sigma,\theta)V_2^*(\sigma,\theta) +G_2^*(\sigma,\theta)V_2(\sigma,\theta)\bigg)+\\
V_2(S,\theta)\int^S d{\sigma\over-2}\bigg(G_2(\sigma,\theta)V_1^*(\sigma,\theta) +G_2^*(\sigma,\theta)V_1(\sigma,\theta)\bigg).
\end{eqnarray*}
It is easy to see the solution of the equation is bounded with respect to the fast variable $S$. 
\paragraph{The third-order correction term of asymptotic expansion (\ref{increasingAsymptoticSolution}).}
The equation for the third-order term of the WKB-solution is following
\begin{eqnarray*}
i\Omega'\pt_S\Phi_3^{(j)}+\bigg(-\theta+2|U^{(j)}|^2\bigg)\Phi_3^{(j)}+\bigg(U^{(j)}{}^2-1\bigg)(\Phi_3^{(j)})^*= G_3,\\
G_3=-i\pt_\theta\Phi_1^{(j)} -i \pt_S\Phi_1^{(j)}\varphi_0'-|\Phi_1^{(j)}|^2 \Phi_1^{(j)} - \\
\Phi_1^{(j)}{}^*\Phi_2^{(j)}U^{(j)}-\Phi_1^{(j)}\Phi_2^{(j)}{}^*U^{(j)} - \Phi_1^{(j)}\Phi_2^{(j)}U^{(j)}{}^*.
\end{eqnarray*}
\par
Equations (\ref{orthogonalityFormulasPostResonantLayer}) give the differential equations for the parameters $A_0(\theta)$ and  $\varphi_0(\theta)$. One can obtain the equations for the main terms of the asymptotic of $A_0(\theta)$ and $\varphi_0(\theta)$ using the obvious form for the asymptotic expansion of $U^{(j)}$ as $\ve\to0$ 
\begin{eqnarray*}
3 A_0+4(1+\theta)A_0'+O(\ve)=0,\\
-2(1+\theta)^{3/2}\varphi_0'+A_0^2(16+12\theta-4\theta^2+(8+8\theta+3\theta^2)\sqrt{\theta+1})+O(\ve)=0.
\end{eqnarray*}
Using regular asymptotic expansion with respect to $\ve$ we obtain:
\begin{eqnarray*}
A_0(\theta)\sim{A_{0,0}\over(1+\theta)^{3/4}},\\
\varphi_0(\theta)\sim\varphi_{0,0}+{A_{0,0}^2\over2}\bigg(5\theta+{3\over2}\theta^2+ {1\over3}(112-8\theta)\sqrt{1+\theta}+3\ln(1+\theta)\bigg).
\end{eqnarray*}
Here $A_{0,0}$ and $\varphi_{0,0}$ are parameters of the asymptotic solution for equation (\ref{main-po-mu}). The values of the parameters will be obtained by matching of the constructed WKB-expansion and the asymptotic expansion in the Painleve layer.
\par
As a result we write out the asymptotic formula for the solution as $\ve\to0$:
\begin{eqnarray*}
\Phi^{(j)}(S,\theta,\ve)\sim (-1)^j\bigg[{1\over2}\sqrt{1+\theta} + \ve^{1/2} A_{0,0}\bigg({1\over\sqrt[4]{1+\theta}}\cos(S) + i\sqrt[4]{1+\theta}\sin(S)\bigg)\bigg],\\
S\sim{\Omega\over\ve}+\varphi_{0,0}+{A_{0,0}^2\over2}\bigg(5\theta+{3\over2}\theta^2+{1\over3}(112-8\theta)\sqrt{1+\theta}+3\ln(1+\theta)\bigg).
\end{eqnarray*}
Here $A_{0,0}$ and $\varphi_{0,0}$ are the parameters of the constructed WKB-solution. 

\paragraph{The domain of validity for the WKB-asymptotic solution at short distance of the slowly varying equilibrium.}
\par
The constructed WKB-asymptotic solution is not valid at the turning points. The higher-order terms of the asymptotic expansion are infinite at these points. The turning points are $\theta=-1$ and $\theta=\infty$. Using obvious formulas for the higher-order terms we reduce:
$$
\Phi_n^{(j)}=O((\theta+1)^{3/2}\Phi_{n-1}^{(j)}), \quad \theta\to-1+0,
$$
Then the domain of validity for the WKB-solution is 
$$
\ve^{-2/3}(1+\theta)\gg1, \quad \theta\to1+0.
$$
When $\theta\to\infty$  the domain of validity for the WKB-solution is obtained analogously:
$$
\Phi_n^{(j)}=O(\theta^{5/4}\Phi_{n-1}^{(j)}), \quad \theta\to\infty,
$$
therefore:
$$
\theta\ll\ve^{-4/5},\quad \theta\to\infty.
$$
\paragraph{Matching the expansion in the Painleve layer and the captured WKB-asymptotic expansion}
\par
Asymptotic expansion (\ref{asymptoticSolutionOfPerturbedPainleveEquation}) is valid as \newline$-\ve^{2/3} \eta \ll1$ or $(\theta+1)\ll1$ for $\theta$ close to $-1$. The captured asymptotic expansion is valid as $\ve^{-2/3}(1+\theta)\gg1$. Therefore captured asymptotic solution (\ref{increasingAsymptoticSolution}) and asymptotic expansion (\ref{asymptoticSolutionOfPerturbedPainleveEquation}) are valid both in the domain $ \ve^{2/3}\ll(1+\theta)\ll1$. Using the uniqueness theorem for the asymptotic expansions one can obtain that these  expansions coincide in this domain. The usual matching procedure \cite{Ilin}  gives us the connection formulas for the parameters $A_k$ and $\phi_k$ of expansion (\ref{increasingAsymptoticSolution}) and parameters of expansion (\ref{asymptoticSolutionOfPerturbedPainleveEquation}). For example, the parameters of the first correction term of (\ref{increasingAsymptoticSolution}) are:
\bb
\varphi_{0,0}=\upsilon,\quad A_{0,0}=\rho.\label{matchingAfterTheCapture}
\ee
\par
Formulas (\ref{matchingBeforeTheCapture}), (\ref{matchingAfterTheCapture}) and theorem \ref{theoremByItsKapaevBelogrudov} give us the statement of theorem \ref{mainStatement}.

\par
{\bf Acknowledgments.} We are grateful to  L. Friedland, R.N. Garifullin, L.A. Kalyakin and B.I. Suleimanov for helpful comments and discussions.

\end{document}